\def\isArxivMode{1}%
\newif\ifarxiv

\ifdefined\isArxivMode
  \arxivtrue
\else
  \arxivfalse
\fi

\documentclass[twocolumn,english,aps,prl,superscriptaddress,nofootinbib]{revtex4-2}%
\usepackage{amsfonts}
\usepackage[T1]{fontenc}
\usepackage{amsmath}
\usepackage{amssymb}
\usepackage{babel}
\usepackage{lastpage}
\usepackage{xcolor}
\usepackage[hidelinks]{hyperref}
\usepackage{braket}

\usepackage{graphicx}
\graphicspath{{figures/}}

\ifarxiv
\else
	\usepackage{xr-hyper}
\fi

\begin{document}
	
	\title{Sensing relativistic quantum fields with minimally perturbing local measurements}
	
	\author{F. Daem}
	
	\affiliation{Laboratoire de Physique Th\'eorique et Mod\'elisation, CNRS Unit\'e 8089, CY
		Cergy Paris Universit\'e, 95302 Cergy-Pontoise cedex, France}
	
	\author{L. Ballesteros Ferraz}
	
	\affiliation{Laboratoire de Physique Th\'eorique et Mod\'elisation, CNRS Unit\'e 8089, CY
		Cergy Paris Universit\'e, 95302 Cergy-Pontoise cedex, France}
	
	\author{A. Zampeli}
	
	\affiliation{International Center for Theory of Quantum Technologies,
		University of Gdansk, Wita Stwosza 63, 80-308 Gdansk, Poland}
	
	\author{A. Matzkin}
	
	\affiliation{Laboratoire de Physique Th\'eorique et Mod\'elisation, CNRS Unit\'e 8089, CY
		Cergy Paris Universit\'e, 95302 Cergy-Pontoise cedex, France}

	\begin{abstract}
		We develop a framework for minimally perturbing local measurements in
		relativistic  quantum field theory,
		with the aim to sense local properties of the field in a non-destructive manner. 
		The field properties are sensed by weakly coupled pointers
		and encapsulated in conditional expectation values
		dependent on a postselection of the field state.
		Our operational protocol uses causally admissible Kraus updates for the field,
		in line with recent relativistic measurement theories,
		keeping in mind restrictions related to ``impossible measurements''.
		We illustrate our approach with three applications:
		a spacelikeness detector for causal-structure sensing,
		counting particle-creation densities in a supercritical potential and
		non-destructive discrimination between entangled states of the field and mixtures.
	\end{abstract}
	
	\maketitle

\textit{Introduction}---Acquiring information on the properties of
relativistic quantum fields involves measuring the fields. As it has been
well-known since the early days of relativistic quantum theory~\cite{landauErweiterungUnbestimmtheitsprinzipsFuer1931,bohrFieldChargeMeasurements1950}, the usual prescriptions of
standard non-relativistic quantum mechanics (NRQM) to account for measurements,
such as projective measurements, lead to ``\emph{absurd results}''~\cite{landauErweiterungUnbestimmtheitsprinzipsFuer1931}.
Local measurements differ from quantum field theory (QFT) predictions in scattering experiments 
that rely on
asymptotic amplitudes, not on local interactions involving field observables.
Today, there is still no consensus on a local measurement theory for
relativistic QFT~\cite{hellwigFormalDescriptionMeasurements1970,aharonovStatesObservablesRelativistic1980,peresQuantumInformationRelativity2004,benincasaQuantumInformationProcessing2014,martin-martinezCausalityIssuesParticle2015,fewsterMeasurementQuantumField2025}.

A widely used approach, pioneered by Unruh and DeWitt~\cite{unruhNotesBlackholeEvaporation1976,dewittQuantumGravityNew1979}
consists in coupling the quantum field to a local detector: the field
properties are then retrieved by measuring the detector. These approaches have
recently gained much traction, partly in order to construct a toolbox for
protocols in relativistic quantum information~\cite{huRelativisticQuantumInformation2012,leeSpatiallyExtendedUnruhDeWitt2014,simidzijaTransmissionQuantumInformation2020,anastopoulosQuantumFieldTheory2023a}, partly to
investigate fundamental problems in relativistic QFT
such as Sorkin's ``impossible measurements'' scenarios~\cite{sorkinImpossibleMeasurementsQuantum1993a,borstenImpossibleMeasurementsRevisited2021,bostelmannImpossibleMeasurementsRequire2021,papageorgiouEliminatingImpossibleRecent2024,oecklCausalMeasurementQuantum2026,oecklLocalCompositionalMeasurements2025}, whose solution hinges on formulating
an adequate measurement theory for QFT. The typology of physically admissible detector models
is currently being investigated~\cite{deramonRelativisticCausalityParticle2021,bednorzGeneralQuantumMeasurements2023,percheParticleDetectorsLocalized2024b,simmonsFactorisationConditionsCausality2025}.

In this paper we build on the local detector models approach in order to
propose a scheme to account for non-destructive measurements of a quantum
field. The main idea is to weakly couple several localized pointers to the
field in different regions of Minkowski spacetime, resulting in entanglement
between the field and the pointers. The field is then postselected by
interacting with detectors on a final region, breaking the entanglement
and leaving the quantum state of each weakly coupled pointer in a modified
state that is finally measured in order to extract the information on the
property of the field.

Here we will restrict our attention mostly to a charged scalar field in flat spacetime. We will first
introduce the weak coupling scheme with postselection, and illustrate our
results with three topical examples of interest to different practitioners of
QFT. In a first example,
we will introduce an array of sensors whose quantum state is
modified by a local transformation of the field in a remote region, indicating
whether each sensor lies inside or outside the forward lightcone of that
region. A second example will deal with the detection of pair-creation by a
supercritical potential, whereby the weakly coupled pointers encode dynamical
properties of the pair-creation process, such as the local particle number
density. A third example will involve discriminating entangled states
from classical mixtures non-destructively for a multi-particle field state.

\begin{figure}[]
\centering
\includegraphics[]{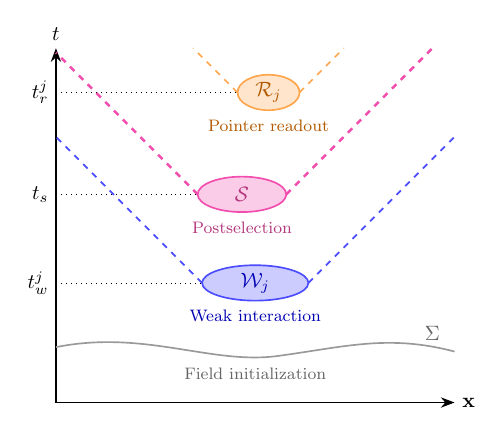}
\caption{Spacetime geometry of the protocol.
The quantum field state $\ket{\psi}$ is prepared on the initial hypersurface $\Sigma$.
The weak interaction between the field and the $j$th pointer occurs in $\mathcal{W}_{j}$ around time $t_{w}^{j}$,
postselection in $\mathcal{S}$ around $t_{s}$,
and the $j$th pointer readout in $\mathcal{R}_{j}$ around $t_{r}^{j}$.}
\label{fig:general}
\end{figure}

\textit{Basic scheme}---We implement non-destructive field measurements in the
following way (see Fig.~\ref{fig:general}). The field is first prepared in
an initial state $\ket{\psi}$ on an initial hypersurface
$\Sigma$. We define one or several regions of compact support in the causal
future $J^{+}(\Sigma)$ of $\Sigma$. In each of these regions $\mathcal{W}_{j}$, $j=1,2,...$,
we weakly couple a field observable to a dynamical observable
$\hat{P}_{j}$ of a non-relativistic pointer that has been prepared in a state
$\Ket{\zeta_{j}^{0}}$. This is a weak unitary coupling
implemented by an interaction Hamiltonian density of the form
\begin{equation}
\hat{\mathcal{H}}_{j}(x)=\lambda\Lambda_{\mathcal{W}_{j}}(x)
\hat{P}_{j}(x)\hat{O}_{j}(x)\,,
\label{int-ham}
\end{equation}
where $\lambda$ is a small coupling constant,
$\Lambda_{\mathcal{W}_{j}}(x)=\omega_{j}(t)f_{j}(\mathbf{x})$ is a smearing function localized around
the time $t_{w}^{j}$ and spatially supported within $\mathcal{W}_{j}$. $\hat{O}_{j}(x)$
is a local observable of the field, written in terms of the field operator in $n$ spatial dimensions
\begin{equation}
\hat{\Phi}(t,\mathbf{x})=\int\mathrm{d}^{n}\mathbf{p}\,
\left(\hat{b}_{\mathbf{p}}(t)u_{\mathbf{p}}(\mathbf{x})
+\hat{d}_{\mathbf{p}}^{\dagger}(t)w_{\mathbf{p}}(\mathbf{x})\right)
\label{phit}
\end{equation}
where $\hat{b}_{\mathbf{p}}(t)$ and $\hat{d}_{\mathbf{p}}(t)$ are the annihilation operators for
particles and antiparticles resp., obeying the equal-time commutation (anticommutation) relations
$[\hat{b}_{\mathbf{p}},\hat{b}_{\mathbf{p}^{\prime}}^{\dagger}]_{\mp}
=[\hat{d}_{\mathbf{p}},\hat{d}_{\mathbf{p}^{\prime}}^{\dagger}]_{\mp}
=\delta^n(\mathbf{p}-\mathbf{p}^\prime)$;
$u_{\mathbf{p}}(x)$ and $w_{\mathbf{p}}(x)$ are the positive
and negative energy solutions
$\left(\left\vert E_{\mathbf{p}}\right\vert =\sqrt{\mathbf{p}^{2}c^{2}+m^{2}c^{4}}\right)$
obeying the free Klein-Gordon (1D Dirac) equation 
(from now on we will set units with $\hbar=c=m=1$).
The unitary evolution $\hat{U}_{j}$ corresponding to $\hat{\mathcal{H}}_{j}$
is taken to first order in $\lambda$ (see Appendix~\ref{app-sec:weak-coupling}).
After the weak interactions are turned off, the field, entangled with
the weakly coupled pointers, evolves freely, until it is postselected in a
state $\ket{\chi}$ (which might still be equal to $\ket{\psi}$ in general) in a region $\mathcal{S}$ in the causal
future of $\Sigma$ and of the regions $\mathcal{W}_{j}$. The postselection
process is implemented by coupling the field to a non-relativistic detector
with dynamical variable $\hat{P}_{s}$ and the interaction Hamiltonian is taken to be
similar to Eq.~\eqref{int-ham}, $g\Gamma_{\mathcal{S}}(x)\hat{P}_{s}(x)\hat{O}_{s}(x)$ except that
here the coupling constant $g$ does not need to be weak. The detector is then measured, and
postselection consists in filtering the final state of the detector. It can be
shown (see App.~\ref{app-sec:kraus}) that the field is then in state
$\ket{\chi} = \hat{K}_{\mathcal{S}}\ket{\psi} $ where $\hat{K}_{\mathcal{S}}$ is a Kraus
operator that can be approximated as
\begin{equation}
\hat{K}_{\mathcal{S}}=\alpha I+
\beta\int \mathrm{d}^{n+1} x^{\prime}\,\Gamma_{\mathcal{S}}(x^{\prime})
\hat{O}_{s}(x^{\prime}) \,;
\label{kraus}%
\end{equation}
$\alpha$ and $\beta$ are numbers depending on the initial and postselected detector states, $\hat{O}_{s}$ is the field observable, and $\Gamma_{\mathcal{S}}$ is a smearing function supported within $\mathcal{S}$.

The last step consists in measuring the weakly coupled pointers after the
field has been postselected. Indeed, the state $\Ket{\zeta_{j}^{0}}$ becomes after postselection
$\Ket{\zeta_{j}^{s}}
= {\bra{\chi} \hat{U}_{j} \ket{\psi}} \Ket{\zeta_{j}^{0}}$
which simplifies
(see App.~\ref{app-sec:pointer-state}) to
\begin{equation}
\exp\left(-i\lambda
\int \mathrm{d}^{n+1} x\,  \Lambda_{\mathcal{W}_{j}}(x)
\frac{\bra{\chi} \hat{O}_j(x)\ket{\psi}}
{\braket{\chi|\psi}}
\hat{P}_{j}(x) \right)
\Ket{\zeta_{j}^{0}} \,;
\label{shift-1}%
\end{equation}
hence after postselection, the state of each weakly coupled pointer is
modified by a quantity depending on
$\frac{\bra{\chi}\hat{O}_j(x)\ket{\psi}}{\braket{\chi|\psi}}$
which is reminiscent of
$A^{w}\equiv\frac{\bra{\eta} \hat{A}\ket{\varphi}}{\braket{\eta|\varphi}}$
known in NRQM as the weak value of an observable $\hat{A}$ of a
system preselected in state $\ket{\varphi}$ and
postselected in state $\ket{\eta}$~\cite{aharonovHowResultMeasurement1988a}.
In NRQM, $\ket{\eta}$ is the eigenstate of an observable. Weak
values have been employed in theoretical and experimental works
in precision metrology~\cite{hostenObservationSpinHall2008,kimHeisenbergLimitedMetrologyWeakValue2022},
state tomography~\cite{lundeenDirectMeasurementQuantum2011,thekkadathDirectMeasurementDensity2016},
or fundamentals of measurements~\cite{panWeaktostrongTransitionQuantum2020,denkmayrObservationQuantumCheshire2014,ballesterosferrazRelevanceWeakMeasurements2024}.
Weak values have been extended to classical field contexts~\cite{dresselClassicalFieldApproach2014a},
path integrals~\cite{turokQuantumTunnelingReal2014,georgievProbingFiniteCoarsegrained2018,matzkinWeakValuesPath2020}, curved spacetime~\cite{fooGeneralrelativisticParticleTrajectories2025}, and have
been used in QFT to formalize the response to a perturbation~\cite{broutQuantumSourceBack1995a,englertHiddenHorizonBlack2010a}. 

Here, the local character of the coupling forces the weak-value
like quantity to be averaged over the spacetime interaction region
$\mathcal{W}_{j},$ weighted by the smearing function
$\Lambda_{\mathcal{W}_{j}}$.
Note that if the detector couples uniformly over the region
$\mathcal{W}_{j}$ Eq.~\eqref{shift-1} simplifies to
$\Ket{\zeta_{j}^{s}} = \exp\left(  -i\lambda
O_{j}^{w}\hat{P}_j\right) \Ket{\zeta_{j}^{0}}$ with
\begin{equation}
O_{j}^{w}=\int \mathrm{d}^{n+1} x \,\Lambda_{\mathcal{W}_j}(x)
\frac{\bra{\psi}\hat{K}_{\mathcal{S}}^{\dagger}\hat{O}_j(x)\ket{\psi}}
{\bra{\psi} \hat{K}_{\mathcal{S}}^{\dagger}\ket{\psi}}
\label{wv1}%
\end{equation}
that will be called here the \textquotedblleft effective weak
value\textquotedblright. Indeed, the pointer state takes the same form as it would in NRQM,
replacing the weak value $A^{w}$ by $O_{j}^{w}$.
If $\Ket{\zeta_{j}^{0}}$
is a Gaussian distribution and $\hat{P}_{j}$ the pointer momentum, the mean position of the
pointer after postselection $\Ket{\zeta_{j}^{s}}$ is shifted by
$\lambda\operatorname{Re}(O_{j}^{w})$~\cite{jozsaComplexWeakValues2007}.
$O_{j}^{w}$ gives information on the local value
of the field observable $\hat{O}_{j}(x)$ in the spacetime region
$\mathcal{W}_{j}$, conditioned on a selective field
measurement in region $\mathcal{S}$ (which may be parsed into several
subregions for a multiparticle field state, as illustrated below). To first order, the weakly coupled pointers are uncorrelated,
enabling simultaneous lightcone
mapping in one run without any update of the field state at intermediate times $t_{w}^{j}$. We now illustrate the formalism with examples relevant to
different uses of QFT.

\textit{Spacelikeness and kick detection}---We investigate here
how a
unitary kick of the field in a region $\mathcal{K}$ of compact spatial support can be
non-destructively detected by weakly coupled pointers placed in different regions. This example is related to the \textquotedblleft
impossible measurements\textquotedblright\ scenario put forward by Sorkin~\cite{sorkinImpossibleMeasurementsQuantum1993a},
that has recently seen a renewed interest~\cite{borstenImpossibleMeasurementsRevisited2021,bostelmannImpossibleMeasurementsRequire2021,papageorgiouEliminatingImpossibleRecent2024,oecklCausalMeasurementQuantum2026,oecklLocalCompositionalMeasurements2025}. 
The scenario is characterized by a kick that can be detected in
a spacelike region and therefore give rise to signaling when a
projective measurement is made in an intermediate region.
Our illustration evades signaling by relying on local and weak unitary interactions with non-relativistic pointers~\cite{deramonRelativisticCausalityParticle2021}.

\begin{figure}[]
\centering
\includegraphics[]{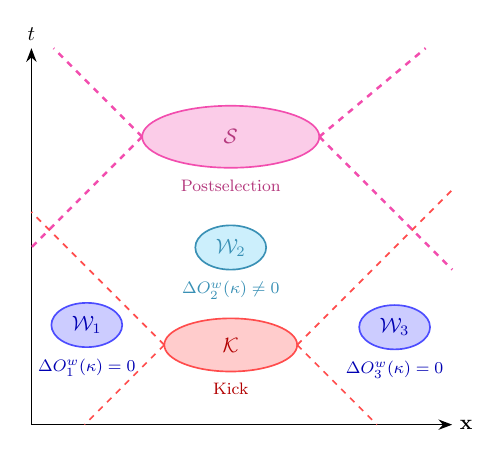}
\caption{Non destructive causal mapping of a local change of the field. A kick in region $\mathcal{K}$
is probed by weakly coupled pointers in regions $\mathcal{W}_{j}$.
Our protocol shows, using the difference $\Delta O^{w}_{j}(\kappa)=O^{w}_{j}(\kappa)-O^{w}_{j}(\kappa=0)$,
that the shift of the $j$th pointer after postselection in $\mathcal{S}$ has no dependence on $\kappa$
($\Delta O^{w}_{j}(\kappa)=0$) if $\mathcal{W}_{j}$ (blue) is outside the lightcone of $\mathcal{K}$,
but is $\kappa$-dependent ($\Delta O^{w}_{j}(\kappa)\neq 0$) if $\mathcal{W}_{j}$ (cyan) is causally connected to $\mathcal{K}$.}%
\label{fig:spacelikeness}%
\end{figure}

Consider a field prepared in a state $\ket{\psi}$.
A ``kick'' given by the unitary
\begin{equation}
\hat{U}_{\kappa}=\exp\left(  -i\kappa\int\mathrm{d}^{n+1}y\,
\Lambda_{\mathcal{K}}(y)\hat{X}(y)\right)  \,;
\label{eq:unitary-kick}
\end{equation}
is implemented in a region $\mathcal{K}$ (see Fig.~\ref{fig:spacelikeness}). We weakly couple
the field to pointers placed in different regions $\mathcal{W}_{j}$ inside or outside the forward lightcone $J^{+}(\mathcal{K})$; each 
pointer is endowed with a dynamical variable $\hat{P}_{j}$ that couples to the field. After postselection 
in  a region $\mathcal{S}$, some pointer states will have shifted by a $\kappa$-dependent effective weak value $O_{j}^{w}(\kappa)$. 

The kick is chosen to be generated by 
$\hat{X}(y)=\hat{\Phi}(y)+\hat{\Phi}^{\dagger}(y)$, 
a \textquotedblleft quadrature operator\textquotedblright :
if $\ket{\psi}$ is chosen as a coherent state $\hat
{U}_{\kappa}$ appears as a field displacement operator (see App.~\ref{app-sec:ex1-coherent}).
For definiteness we will take the weakly coupled  field observable in each $\mathcal{W}_j$ 
to be the normal ordered number density
$\hat{O}_{j}(x) = {:\!\hat{\Phi}^{\dagger}(x)\hat{\Phi}(x)\!:}$.
Concerning postselection, the field observable must be invariant relative to the kick:
otherwise the correlations would systematically depend on
$\hat{U}_{\kappa}$ through the local values of the field in
$\mathcal{S}$. Here, we set
$\hat{O}_{s}(x)=i(\hat{\Phi}(x)-\hat{\Phi}^{\dagger}(x))$.
It can indeed be checked that with this choice
$[\hat{K}_{\mathcal{S}},\hat{U}_{\kappa}]=0$
(see App.~\ref{app-sec:ex1-postselection}).

We can now determine the effective weak value (see App.~\ref{app-sec:ex1-wv})
\begin{widetext}
\begin{equation}
	O^{w}_{j}(\kappa) = O^{w}_{j}(\kappa=0)
	+ \int \mathrm{d}^{n+1} x\,\Lambda_{\mathcal{W}_j}(x)
	\left\{
	\kappa \int \mathrm{d}^{n+1} y \, \Lambda_{\mathcal{K}}(y)
	\Delta(x-y) \, X^w(x)
	+\kappa^2 \left(\int\mathrm{d}^{n+1} y\, \Lambda_{\mathcal{K}}(y)
	\Delta(x-y)\right)^2 \right\} \,,
	\label{eq:wv-kicked}
\end{equation}
\end{widetext}
where $\Delta(x-y)$ is the Pauli-Jordan function and $X^w(x)$ is the weak value density of the $\hat{X}$ quadrature.
Given that $\Delta(x-y)$ vanishes for spacelike intervals, Eq.~\eqref{eq:wv-kicked} tells us that 
pointers in regions $\mathcal{W}_j$ outside the causal future of $\mathcal{K}$ will not
have a $\kappa$-dependent shift. $O^{w}_{j}(\kappa)$  depends on the kick only if  $\mathcal{W}_j$ is
in  $J^{+}(\mathcal{K})$. Operationally, the  difference $\Delta O_{j}^{w}(\kappa)=O_{j}^{w}(\kappa)-O_{j}^{w}(\kappa=0)$
does not vanish for such pointers, while those for which $\Delta O_{j}^{w}(\kappa)=0$ are necessarily 
spacelike to the kick region. Note that the non-destructive feature allows for monitoring
of the kick in different regions simultaneously, while a selective measurement in $\mathcal{S}$ can
amplify the $\kappa$-depedence of the effective weak value.

\begin{figure}[]
	\centering
	\includegraphics{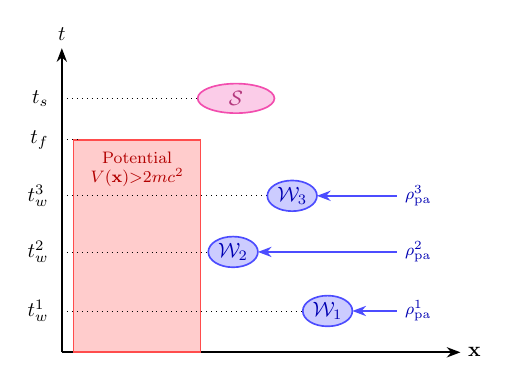} \caption{Schematics of non-destructive 
		detection of the local particle density produced by a supercritical background field ($V(\mathbf{x})>2mc^2$). 
		The space-time diagram shows regions $\mathcal{W}_j$ where 
		weak unitary couplings take place during a short time interval centered on $t_{w}^{j}$.
		The field is postselected in region $\mathcal{S}$ at time $t_{s}$.}
	\label{fig:example2}
\end{figure}

\textit{Pair creation dynamics}---As is well-known~\cite{fedotovAdvancesQEDIntense2023},
a background supercritical field creates particle-antiparticle pairs.
For instance a static potential barrier $V(\mathbf{x})$ of strength $V_{\max}>2mc^{2}$
radiates (see Fig.~\ref{fig:example2}): if the potential is turned on at
$t=0$ when the initial state is the vacuum $\ket{0}$, the
number density of electrons at time $t_{f}$ and position
$\mathbf{x}$ is computed~\cite{chengIntroductoryReviewQuantum2010,alkhateebSpacetimeresolvedQuantumField2022} through the vacuum expectation value
$\rho_{\mathrm{pa}}(t_{f},\mathbf{x})$
of $\hat{\rho}_{\mathrm{pa}}\equiv\hat{\Phi}_{\mathrm{pa}}^{\dagger}\hat{\Phi}_{\mathrm{pa}}$.
$\hat{\Phi}_{\mathrm{pa}}$ is the positive frequency part of Eq.~\eqref{phit},
that now represents a Dirac field in one spatial dimension.
$\hat{\Phi}(x)$ fulfills
$i\partial_{t}\hat{\Phi}=\left(  h_{0}+V\right)  \hat{\Phi}$,
where $h_{0}=-ic\sigma_1 \partial_x+\sigma_2 mc^2$
is the 1D Dirac Hamiltonian (recall spin is frozen in 1D).
Note that at $t_{f}$ the potential must be turned off,
given that otherwise $\ket{0}$ would not represent the correct vacuum~\cite{gongBirthProcessElectronpositron2023}.
We can circumvent this issue and follow the
dynamics of particle creation in time without turning off the field by weakly
coupling pointers in the following way.

In order to monitor non-destructively say the particle density
in a region $\mathcal{W}_{j}$, given by
$\rho_{\mathrm{pa}}^{j} = \int \mathrm{d}^{1+1} x\,\Lambda_{\mathcal{W}_{j}}(x)\rho_{\mathrm{pa}}(x)$,
take $\hat{O}_{j}(x)=\hat{\rho}_{\mathrm{pa}}(x)$ and
consider Eq.~\eqref{wv1}.
Choosing the postselection observable in region $\mathcal{S}$ to be
$\hat{O}_{s}(x)=\hat{\rho}_{\mathrm{pa}}(x)$
and the associated detector to be a 2-level
system, with $\hat{P}_{s}=\sigma_{y}$ and
$\ket{d_{0}}=\ket{{+z}}$ (eigenstate of $\sigma_{z}$),
we find that the
$j$th pointer shift depends on (see App.~\ref{app-sec:ex2})
\begin{equation}
O_{j}^{w}= \rho_{\mathrm{pa}}^{j}
+ig {\bra{d_{s}} \sigma_{y} \ket{{+z}}}^{\ast}
C_{j,\mathcal{S}}\,,
\label{rho-weak}%
\end{equation}
\begin{widetext}
\begin{equation}
C_{j,\mathcal{S}}=
\frac{\displaystyle\int \mathrm{d} \mathbf{p}\,\mathrm{d} \mathbf{p}^{\prime}\,
\int \mathrm{d}^{1+1} x\,\Lambda_{\mathcal{W}_j}(x)
\left(\tilde{u}_{\mathbf{p}}^{(+)}(x)\right)^\dagger
\tilde{w}_{\mathbf{p}^{\prime}}^{(+)}(x)
\int \mathrm{d}^{1+1} x^{\prime}\,\Gamma_{\mathcal{S}}(x^{\prime})
\left[\left(\tilde{u}_{\mathbf{p}}^{(+)}(x^{\prime})\right)^{\dagger}
\tilde{w}_{\mathbf{p}^{\prime}}^{(+)}(x^{\prime})\right]^{\ast}}
{\displaystyle\braket{d_{s}|{+z}}^{\ast}
+ig {\bra{d_{s}}\sigma_{y} \ket{{+z}}}^{\ast}
\int \mathrm{d}^{1+1} x^{\prime}\,\Gamma_{\mathcal{S}}(x^{\prime})
\rho_{\mathrm{pa}}(x^{\prime})}\,,
\label{rho-weak_Tkf}
\end{equation}
\end{widetext}
where $\tilde{u}_{\mathbf{p}}^{(+)}(t,\mathbf{x})$
($\tilde{w}_{\mathbf{p}}^{(+)}(t,\mathbf{x})$) denote
the positive (negative) sector of the basis states of the
expansion Eq.~\eqref{phit} (see App.~\ref{app-sec:ex2}).
We have assumed $\braket{d_{s}|{+z}} \neq 0$ and
to simplify the expression, that the field dynamics is not modified during the
time of the weak coupling, whose effective time is $t_{w}^{j}$.
By choosing the postselected detector state in $\mathcal{S}$ to be
$\ket{{+z}}$,
the second term in Eq.~\eqref{rho-weak} vanishes and
$O_{j}^{w}=\rho_{\mathrm{pa}}^{j}$: by placing several weakly coupled
pointers in different regions $\mathcal{W}_j$, one can sense the local spacetime
density of the created particles.
The term $ C_{j,\mathcal{S}}$ encapsulates the
non-factorizable part of the correlators
${\bra{0}\hat{O}_{s}(x^{\prime})\hat{O}_{j}(x)\ket{0}}$.
This involves the field correlations between the regions $\mathcal{S}$ and $\mathcal{W}_j$
and is quantified by the values of the observable sensed in each region.
Expressions similar to Eq.~\eqref{rho-weak} hold for other observables coupled
in region $\mathcal{W}_{j}$; for instance for the total charge density
$\hat{O}_{j}(x)=\hat{Q}(x)\equiv {:\!\hat{\Phi}^{\dagger}(x)\hat{\Phi}(x)\!:}$.

\begin{figure}[]
	\centering
	\includegraphics[]{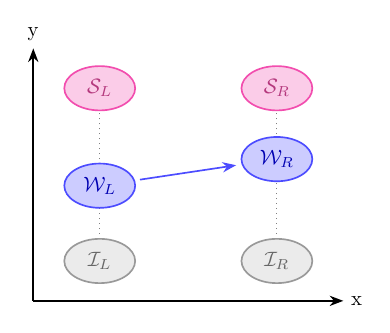} 
	\caption{2D spatial setup ($\mathbf{x}=(\mathrm{x},\mathrm{y})$) for discriminating an entangled state from a classical
		mixture. A 2-particle field state (entangled or mixture) is prepared with a
		particle in region $\mathcal{I}_{L}$, another in region $\mathcal{I}_{R}$. In region $\mathcal{W}_{L}$, the field is weakly
		coupled to a pointer. The pointer is moved to region $\mathcal{W}_{R}$ and coupled there to
		the field. Finally the field is postselected in regions $\mathcal{S}_{L}$ and $\mathcal{S}_{R}$. }
	\label{fig:example3}
\end{figure}

\textit{Sensing entanglement}---Local-probe
approaches to quantum-field entanglement usually aim to extract correlations between spacetime
regions~\cite{huRelativisticQuantumInformation2012,reznikViolatingBellsInequalities2005,des.l.torresEntanglementStructureQuantum2023a}.
Here we propose instead a protocol sensing entangled states of a field.
Consider two distinct regions
$\mathcal{I}_{L}$ and $\mathcal{I}_{R}$ on
the same hypersurface $\Sigma$ in the lab frame and define the bosonic two-particle state
$\ket{\varphi \,\eta}=\hat{b}^{\dagger}(\varphi)\,
\hat{b}^{\dagger}(\eta)\ket{0}$
where
$\hat{b}^{\dagger}(\varphi)=\int\mathrm{d}^{n}\mathbf{p}\,
\varphi(\mathbf{p})\,\hat{b}_{\mathbf{p}}^{\dagger}$ creates a localized
wavepacket (e.g.\ with Gaussian profiles). Assume we can prepare an entangled state,
$\ket{\psi_{\mathrm{ent}}}
=c_{1} \ket{\varphi_{L}\eta_{R}}
+c_{2} \ket{\eta_{L}\varphi_{R}}$ or a mixture with identical diagonal elements,
$\rho_{\mathrm{mix}}
=|c_{1}|^{2} {\ket{\varphi_{L}\eta_{R}} \bra{\varphi_{L}\eta_{R}}}
+|c_{2}|^{2} {\ket{\eta_{L}\varphi_{R}} \bra{\eta_{L}\varphi_{R}}}$.
We show, using the specific 2D geometry
pictured in Fig.~\ref{fig:example3} how weak couplings can discriminate
non-destructively $\ket{\psi_{\mathrm{ent}}}$ from $\rho_{\mathrm{mix}}$.

Consider regions $\mathcal{W}_{j}$, $j=R,L$ (Fig.~\ref{fig:example3}). Let us assume a weak
unitary coupling is implemented in region $\mathcal{W}_{L}$ between the field and a
non-relativistic pointer, at time $t_{w}^{L}$ when the wavepacket is in region $\mathcal{W}_{L}$.
We then couple the same pointer in region $\mathcal{W}_{R}$ at time $t_{w}^{R}$ when the
wavepacket passes by region $\mathcal{W}_{R}$. Finally postselection takes places at $t_{s}$
in two distinct subregions $\mathcal{S}_{L}$ and $\mathcal{S}_{R}$ defined over the
hypersurface $\Sigma_{S}$. In each subregion a detector couples to the number-density observable
$\hat{\rho}(x)\equiv {:\!\hat{\Phi}^{\dagger}(x)\hat{\Phi}(x)\!:}$.
In regions $\mathcal{W}_{j}$ we also
couple $\hat{O}_{j}(x)=\hat{\rho}(x)$ to the weak pointer.

The Kraus operator becomes $\hat{K}_{\mathcal{S}}=\hat{K}_{\mathcal{S}_{L}}\hat{K}_{\mathcal{S}_{R}}$
where each $\hat{K}_{\mathcal{S}_{l}}$, $l=L,R$, is given by Eq.~\eqref{kraus} (see App.~\ref{app-sec:ex3-kraus}).
Given that we have sequential interactions the weak pointer shifts due to the couplings in regions $\mathcal{W}_{L}$
and $\mathcal{W}_{R}$ add up and
the total pointer shift that can be read on the pointer at $t>t_{s}$ is
proportional to $\sum_{j}\operatorname{Re}(O_{j}^{w})$.
For a pure state, the effective weak value $O_{j}^{w}$ takes the
form given by Eq.~\eqref{wv1}, while the generalization for a density matrix $\rho_{\mathrm{mix}}$
takes the form (see App.~\ref{app-sec:ex3-WV})
\begin{equation}
	O_{j}^{w}=\int \mathrm{d}^{n+1} x\,\Lambda_{\mathcal{W}_j}(x)
	\frac{\mathrm{Tr}_{\Phi}
	\left(\rho_{\mathrm{mix}} {\hat{K}_{\mathcal{S}}^{\dagger }\hat{K}_{\mathcal{S}}^{}}\hat{O}_{j}(x)\right)}
	{\mathrm{Tr}_{\Phi}
	\left(\rho_{\mathrm{mix}} {\hat{K}_{\mathcal{S}}^{\dagger }\hat{K}_{\mathcal{S}}^{}}\right)}\,.
	\label{eq:density-matrix-wv}
\end{equation}
We show (see App.~\ref{app-sec:ex3-determination-pointer-shift}) that the difference of pointer shifts $\lambda\sum_{j}\operatorname{Re}(O_{j}^{w})$
between entangled states and a mixture
lies in the presence of the nondiagonal terms of the form
${\bra{\varphi_{L}\eta _{R}}\hat{K}_{\mathcal{S}}^{\dagger }\hat{K}_{\mathcal{S}}^{}
\hat{O}_{j}(x) \ket{\eta_{L}\varphi_{R}}}$ and
${\bra{\varphi_{L}\eta_{R}} \hat{K}_{\mathcal{S}}^{\dagger}\hat{K}_{\mathcal{S}}^{}
\ket{\eta_{L}\varphi_{R}}}$
in the entangled case.
These are generically nonzero, leading to different pointer statistics.

\textit{Conclusion}---We have introduced a framework accounting for minimally
perturbing non-destructive local measurements in relativistic QFT. Our
framework is based on implementing local weak unitary couplings with
non-relativistic pointers followed by postselection through a Kraus update.
The aim is to harvest information from the field in a non-destructive way, allowing
for further manipulation and control of the field.
The main objects are effective weak values that determine the pointers
shift. Weak values encode local field information: we have
seen that they vanish when the coupling is spacelike to a perturbation and
that they could be useful in pair-production diagnostics. We
have also seen how they can detect entangled states of quantum fields, a first step in
constructing an entanglement witness. We can further
anticipate that our nondestructive local measurement scheme will be useful in
a toolbox for relativistic quantum information.

\begin{acknowledgments}
\textit{Acknowledgments}---This research was supported by the EUTOPIA Science and Innovation Fellowship Programme and funded by the European Union Horizon 2020 programme under the Marie Skłodowska-Curie grant agreement No 945380.
A.Z. acknowledges support by the COST Action CA23115 and the Maestro project no. 2021/42/A/ST2/00356 under the title “Relativistic Causality and Information Processing”, funded by the National Science Center, Poland.
\end{acknowledgments}

\ifarxiv
\onecolumngrid
\appendix
\setcounter{secnumdepth}{2}

\section{Weak measurement scheme}

Throughout this article, we use the notations $\mathbf{x}$ for
the $n$-dimensional spatial coordinates and $x=(t,\mathbf{x})$ for the $(n+1)$-dimensional spacetime coordinates.

\subsection{The weak coupling approximation}
\label{app-sec:weak-coupling}

Using the interaction hamiltonian of Eq.~\eqref{int-ham} of the main text,
the associated evolution operator is given in the interaction picture by
\begin{equation}
	\hat{U}_j = T \exp\left(-i \lambda \int \mathrm{d}^{n+1} x \, \Lambda_{\mathcal{W}_j}(x)
	\hat{P}_j(x) \hat{O}_j(x)\right) \,,
	\label{eq:unitary-evolution}
\end{equation}
where $T$ is the time-ordering operator.
The coupling constant $\lambda$ is assumed to be small, and the interaction is assumed to be localized in a spacetime region $\mathcal{W}_j$.
For a well suited normalization of the smearing function $\Lambda_{\mathcal{W}_j}$, we can expand the evolution operator to first order in $\lambda$ as
\begin{equation}
	\hat{U}_j \approx 1 -i 
	\lambda \int \mathrm{d}^{n+1} x \, \Lambda_{\mathcal{W}_j}(x)
	\hat{P}_j(x) \hat{O}_j(x) \,.
	\label{eq:weak-unitary}
\end{equation}

\subsection{Postselection and the Kraus operator}

\label{app-sec:kraus}

We will call $\ket{d_0}$ the initial state of the detector and $\ket{d_s}$ the
state of the detector when postselection is successful.
The normalized postselected state
of the field is then given by
$\ket{\chi_{s}} = p_s^{-1/2}{\bra{d_s} \hat{U}_{\mathcal{S}} \ket{d_0} \ket{\psi}}$
where $p_s$ is the probability of successful postselection and
the unitary evolution is obtained from the detector-field interaction
Hamiltonian proposed in the main text.
We will use the unnormalized states $\ket{\chi} = p_{s}^{1/2} \ket{\chi_{s}}$
throughout as the objects we are considering do not depend on the normalization.
The coupling constant $g$ is not assumed to be small for an ideal measurement,
but for simplicity we still consider the first order in $g$ for $\hat{U}_{\mathcal{S}}$, similar to the weak coupling approximation in Eq.~\eqref{eq:weak-unitary}. The field state after postselection is then
\begin{equation}
	\ket{\chi} =  \left(\braket{d_s | d_0} -ig \int \mathrm{d}^{n+1} x ~ \Gamma_{\mathcal{S}}(x) \bra{d_s} \hat{P}_{s}(x) \ket{d_0} \hat{O}_{s}(x)\right) \ket{\psi}\,.
\end{equation}
The Kraus operator $\hat{K}_{\mathcal{S}}$ is defined by $\hat{K}_{\mathcal{S}} = {\bra{d_s} \hat{U}_{\mathcal{S}} \ket{d_0}}$, so we can clearly write it in the form of Eq.~\eqref{kraus} of the main text
\begin{equation}
\hat{K}_{\mathcal{S}}=\alpha I+\beta\int \mathrm{d}^{n+1} x^{\prime}\, \Gamma_{\mathcal{S}}(x^{\prime})\hat{O}_{s}(t^{\prime},\mathbf{x}^{\prime})\,,%
\end{equation}
where the detector variable is assumed to couple uniformly over the postselection region $\mathcal{S}$ ($\hat{P}_{s}(x) = \hat{P}_{s}$)
and the coefficients are $\alpha = \braket{d_s | d_0}$ and $\beta = -ig {\bra{d_s} \hat{P}_{s} \ket{d_0}}$.

\subsection{Pointer state after postselection}

\label{app-sec:pointer-state}

We want to express the state of a weakly coupled pointer after a weak measurement. The
initial state of the field $\ket{\psi}$ and the pointer $\Ket{\zeta_j^0}$ is evolved in the interaction picture with the weak interaction
$\hat{U}_j$, then the state of the field is postselected on a state
$\ket{\chi}$. The final pointer state is thus $\Ket{\zeta_j^s} = {\bra{\chi}  \hat{U}_j \ket{\psi}} \ket{\zeta_j^0}$
(up to a normalization factor). Using the weak coupling approximation of Eq.~\eqref{eq:weak-unitary}, we can write
the unnormalized pointer state as
\begin{equation}%
\Ket{\zeta_j^s}  = \left(  \braket{\chi|\psi} -i 
	\lambda \int \mathrm{d}^{n+1} x \, \Lambda_{\mathcal{W}_j}(x)
	\hat{P}_j(x) \bra{\chi}\hat{O}_j(x) \ket{\psi}\right) \Ket{\zeta_j^0}\,.
\end{equation}
We can assume that $\ket{\chi}$ is not orthogonal to $\ket{\psi}$ (otherwise the postselection would be impossible to first order in $\lambda$) so that it is possible to factor out $\braket{\chi|\psi}$ and write the final pointer state as
\begin{equation}
	\Ket{\zeta_j^s} = \left(  1 -i 
	\lambda \int \mathrm{d}^{n+1} x \, \Lambda_{\mathcal{W}_j}(x)
	\hat{P}_j(x) \frac{\bra{\chi}\hat{O}_j(x) \ket{\psi}}{\braket{\chi|\psi}} \right)  \Ket{\zeta_j^0}
\end{equation}
and using the weak coupling approximation again, we can write the final pointer state in the form of Eq.~\eqref{shift-1} of the main text
\begin{equation}
\Ket{\zeta_{j}^{s}} = \exp\left(  -i\lambda
\int \mathrm{d}^{n+1} x \, \Lambda_{\mathcal{W}_{j}}(x)
\frac{\bra{\chi} \hat{O}_j(x)\ket{\psi}}
{\braket{\chi|\psi}}
\hat{P}_{j}(x) \right) \Ket{\zeta_{j}^{0}} \,.
\end{equation}

\section{Spacelikeness and kick detection}

\subsection{Coherent states}
\label{app-sec:ex1-coherent}

The initial state of the field can be chosen to be a multicoherent state
$\ket{\alpha,\beta}$.
Using the kick operator $\hat{U}_\kappa$ defined in Eq.~\eqref{eq:unitary-kick} of the main text,
the state of the field after the kick is
\begin{equation}
\ket{\chi}=\hat{U}_\kappa\ket{\alpha,\beta}
=T\exp\left(  -i\int\mathrm{d}^{n+1}x \,\Lambda_{\mathcal{K}}(x)\left(
	\Phi_{\mathrm{cl}}(x)+\Phi_{\mathrm{cl}}^{\ast}(x)
	\right)\right)  \ket{\alpha,\beta}\,,
\end{equation}
where we used the action of the quadrature operator $\hat{X}$ on the multicoherent
state, which yields a c-number term expressed with the classical field configuration
\begin{equation}
	\Phi_{\mathrm{cl}}(x)=\int \mathrm{d}^{3}\mathbf{p} \,
	\left(  \alpha_{\mathbf{p}}e^{-i p\cdot x}
	+\beta_{\mathbf{p}}^{\ast}e^{i p\cdot x}\right)  \,.
\label{eq:cl-field}
\end{equation}

\subsection{Kick-transparent postselection observable}
\label{app-sec:ex1-postselection}

We consider the specific unitary kick generated by $\hat{X}(y)=\hat{\Phi}(y)+\hat{\Phi}^{\dagger}(y)$ and given by equation~\eqref{eq:unitary-kick} of the main text.
We want to find an observable of the field invariant under this kick,
so that the postselected state is unaffected by the kick.

We first derive the transformation of the field operators under $\hat
{U}_{\kappa}$. Using the Baker-Campbell-Hausdorff (BCH) formula, the transformed
field operator is
\begin{equation}
\hat{\Phi}_{\kappa}(x)
\equiv \hat{U}_{\kappa}^{\dagger}\hat{\Phi}(x) \hat{U}_{\kappa}
= \hat{\Phi}(x) +\left[  i\kappa\int\mathrm{d}^{n+1} y \, \Lambda_{\mathcal{K}}(y)(\hat{\Phi}(y)
+ \hat{\Phi}^{\dagger}(y)),\, \hat{\Phi}(x)\right]_{-}
\label{eq:BCH}
\end{equation}
where we can use $[\hat{\Phi}(x), \hat{\Phi}(y)]_{-} = 0$ and the canonical
commutation relation of the complex scalar field to compute the commutator
\begin{equation}
\left[\hat{\Phi}(x),\, \int\mathrm{d}^{4} y \, \Lambda_{\mathcal{K}}(y)
(\hat{\Phi}(y) + \hat{\Phi}^{\dagger}(y))\right]_{-}
= \int\mathrm{d}^{n+1} y \,\Lambda_{\mathcal{K}}(y)
[\hat{\Phi}(x),\, \hat{\Phi}^{\dagger}(y)]_{-}
= i\int\mathrm{d}^{n+1} y \, \Lambda_{\mathcal{K}}(y)\Delta(x-y) \,,
\end{equation}
where $\Delta(x-y)$ is the Pauli-Jordan function.
This is a c-number, which explains why the
higher-order commutators in the BCH expansion of Eq.~\eqref{eq:BCH} vanish.
Similarly,
\begin{equation}
\left[\hat{\Phi}^{\dagger}(x),\, \int\mathrm{d}^{n+1} y ~ \Lambda
_{\mathcal{K}}(y)(\hat{\Phi}(y) + \hat{\Phi}^{\dagger}(y))\right]_{-}
= \int\mathrm{d}^{n+1} y \, \Lambda_{\mathcal{K}}(y)
[\hat{\Phi}^{\dagger}(x),\, \hat{\Phi}(y)]_{-} = i\int\mathrm{d}^{n+1} y ~ \Lambda
_{\mathcal{K}}(y)\Delta(x-y) \,,
\end{equation}
where only $\hat{\Phi}(y)$ contributes in the calculation. This kick
transforms therefore the field operators as
\begin{subequations}
\label{eq:field_shift}%
\begin{equation}
\hat{\Phi}_{\kappa}(x) = \hat{\Phi}(x) + \kappa
\int\mathrm{d}^{n+1} y ~ \Lambda_{\mathcal{K}}(y) \Delta(x-y)
\end{equation}
and
\begin{equation}
\hat{\Phi}^{\dagger}_{\kappa}(x) = \hat{\Phi
}^{\dagger}(x) + \kappa\int\mathrm{d}^{n+1} y ~ \Lambda_{\mathcal{K}%
}(y) \Delta(x-y) \,.
\end{equation}
\end{subequations}

These results are exact to all orders in $\kappa$ and, crucially, both field
components acquire the same shift, which implies that the other quadratures
$\hat{Y}(x) = i(\hat{\Phi}(x) - \hat{\Phi}^{\dagger}(x))$ and $\hat{P}_{Y}(x)
= i(\hat{\Pi}(x)^{\dagger}- \hat{\Pi}(x))$, where $\hat{\Pi}(x)$ is the canonical conjugate field to $\hat{\Phi}(x)$,
are invariant under the kick.
That is, for any value of $\kappa$, $\hat{U}_{\kappa}^{\dagger}\hat{Y}(x) \hat
{U}_{\kappa}= \hat{Y}(x)$ and $\hat{U}_{\kappa}^{\dagger}\hat{P}_{Y}(x)
\hat{U}_{\kappa}=\hat{P}_{Y}(x)$ even if the spacetime point $x$ is causally
connected to the kick region $\mathcal{K}$.
Consequently, postselection based on measurements of the $\hat{Y}$ or
$\hat{P}_{Y}$ quadrature yields a Kraus operator $\hat{K}_{\mathcal{S}}$
that is invariant under the kick: $\hat{U}_{\kappa}^{\dagger}\hat{K}_{\mathcal{S}} \hat{U}_{\kappa}= \hat{K}_{\mathcal{S}}$.
For example, take
\begin{equation}
	\hat{K}_{\mathcal{S}} = T\exp\left(-i \int \mathrm{d}^{n+1} x \, \Lambda_{\mathcal{S}}(x) \hat{Y}(x)\right) \,,
	\label{eq:quadrature-kraus}
\end{equation}
where $\Lambda_{\mathcal{S}}(x)$ is a real smearing function with support in the postselection region $\mathcal{S}$.
This ensures that the postselected state $\ket{\chi} \propto \hat{K}_{\mathcal{S}} \ket{\psi}$ does not depend on the kick.

\subsection{Effective weak value calculation}
\label{app-sec:ex1-wv}

We want to express the effective weak value of the field observable considered in the main text.
We use the normal-ordered number density $\hat{O}_{j}(x) = {:\!\hat{\Phi}^\dagger(x)\hat{\Phi}(x)\!:}$ and
the unitary kick given in Eq.~\eqref{eq:unitary-kick} of the main text.
Using the field transformations derived in App.~\ref{app-sec:ex1-postselection}, the transformed normal-ordered observable can be computed exactly by expanding the product
\begin{equation}
	\begin{split}
		{:\!\hat{\Phi}^\dagger_\kappa(x)\hat{\Phi}_\kappa(x)\!:} &=
		{:\!\left(\hat{\Phi}^\dagger(x) + \kappa \int \mathrm{d}^{n+1} y \, \Lambda_{\mathcal{K}}(y) \Delta(x-y)\right)
		\left(\hat{\Phi}(x) + \kappa \int \mathrm{d}^{n+1} y \, \Lambda_{\mathcal{K}}(y) \Delta(x-y)\right)\!:} \\
		&= {:\!\hat{\Phi}^\dagger(x)\hat{\Phi}(x)\!:}
		+ \kappa \int \mathrm{d}^{n+1} y \, \Lambda_{\mathcal{K}}(y) \Delta(x-y) \left(\hat{\Phi}^\dagger(x)+ \hat{\Phi}(x)\right)
		+ \kappa^2 \left(\int \mathrm{d}^{n+1} y \, \Lambda_{\mathcal{K}}(y) \Delta(x-y)\right)^2 \,.
	\end{split}
\end{equation}
The associated effective weak value readily simplifies to the form of Eq.~\eqref{eq:wv-kicked} of the main text
\begin{equation}
	\begin{split}
		O^{w}_{j}(\kappa) &= O^{w}_{j}(\kappa=0)
		+ \int \mathrm{d}^{n+1} x \, \Lambda_{\mathcal{W}_j}(x) \left\{
		\kappa \int \mathrm{d}^{n+1} y \, \Lambda_{\mathcal{K}}(y) \Delta(x-y) \, X^{w}(x)
		+ \kappa^2 \left(\int \mathrm{d}^{n+1} y \, \Lambda_{\mathcal{K}}(y) \Delta(x-y)\right)^2 \right\} \,,
	\end{split}
\end{equation}
where $X^w(x)$ is the weak value distribution of the $\hat{X}$ quadrature.

Note that for a coherent state $\ket{\alpha,\beta}$ and the Kraus operator $\hat{K}_{\mathcal{S}}$ defined in Eq.~\eqref{eq:quadrature-kraus}
the post-selected state is
\begin{equation}
\ket{\chi}=p_{s}^{-1/2} \hat{K}_{\mathcal{S}}\ket{\alpha,\beta}
=T\exp\left(  -i\int\mathrm{d}^{n+1}x \, \Lambda_{\mathcal{S}}(x)
i\left(  \Phi_{\mathrm{cl}}(x)-\Phi_{\mathrm{cl}}^{\ast}(x)\right)
\right)  \ket{\alpha,\beta}\,,
\end{equation}
and the weak values can therefore be expressed in term of the classical field configuration defined in Eq.~\eqref{eq:cl-field} as
\begin{equation}
	X^{w}(x) = \frac{\bra{\chi} \hat{X}(x) \ket{\alpha,\beta}}
	{\braket{\chi|\alpha,\beta}}
	= \Phi_{\mathrm{cl}}(x) + \Phi^{\ast}_{\mathrm{cl}}(x) \,,
\end{equation}
and
\begin{equation}
	O^{w}_{j}(\kappa=0) = \frac{\bra{\chi} {:\!\hat{\Phi}^\dagger(x)\hat{\Phi}(x)\!:} \ket{\alpha,\beta}}
	{\braket{\chi|\alpha,\beta}}
	= \Phi^{\ast}_{\mathrm{cl}}(x)\Phi_{\mathrm{cl}}(x) \,.
\end{equation}

\section{Non destructive monitoring of pair-creation dynamics}
\label{app-sec:ex2}

We derive in this Appendix Eq.~\eqref{rho-weak} of the main text giving the
effective weak value in the context of monitoring non-destructively the
dynamics of pair-creation in a supercritical potential.

Eq.~\eqref{phit} of the main text represents here the field operator expansion
in which the action of the background supercritical potential is encapsulated~\cite{chengIntroductoryReviewQuantum2010}
in the time-dependence of the creation and annihilation operators
through a Bogoliubov transformation,
\begin{subequations}
	\label{A-ex2-bogo}
	\begin{align}
	\hat{b}_{\mathbf{p}}(t)
	&=\int \mathrm{d}\mathbf{p}^{\prime}\,\left(
	U_{u_{\mathbf{p}}u_{\mathbf{p}^{\prime}}}(t,t_{0})\,\hat{b}_{\mathbf{p}^{\prime}}(t_{0})
	+U_{u_{\mathbf{p}}w_{\mathbf{p}^{\prime}}}(t,t_{0})\,\hat{d}_{\mathbf{p}^{\prime}}^{\dagger}(t_{0})\right) \,,\\
	\hat{d}_{\mathbf{p}}^{\dagger}(t)
	&=\int \mathrm{d}\mathbf{p}^{\prime}\,\left(
	U_{w_{\mathbf{p}}u_{\mathbf{p}^{\prime}}}(t,t_{0})\,\hat{b}_{\mathbf{p}^{\prime}}(t_{0})
	+U_{w_{\mathbf{p}}w_{\mathbf{p}^{\prime}}}(t,t_{0})\,\hat{d}_{\mathbf{p}^{\prime}}^{\dagger}(t_{0})\right)\,.
	\end{align}
\end{subequations}
$U_{\alpha\beta}$ are the matrix elements of the unitary operator $\hat{U}(t,t_{0})$ generated by
the full Hamiltonian $c\sigma_1 \hat{\mathbf{p}} +\sigma_2 mc^2+V(\mathbf{x})$.
From now on we omit the time argument at $t_0$ and write $\hat{b}_{\mathbf{p}}\equiv \hat{b}_{\mathbf{p}}(t_0)$ and $\hat{d}_{\mathbf{p}}\equiv \hat{d}_{\mathbf{p}}(t_0)$.
In one spatial dimension the spin is frozen (hence no spin index) and $u_{\mathbf{p}}$ and $w_{\mathbf{p}}$ are two-component Dirac
spinors. The creation and annihilation operators obey
\begin{equation}
    \left[\hat{b}_{\mathbf{p}},\,\hat{b}_{\mathbf{q}}^{\dagger}\right]_{+}=\delta(\mathbf{p}-\mathbf{q}),
    \qquad
    \left[\hat{d}_{\mathbf{p}},\,\hat{d}_{\mathbf{q}}^{\dagger}\right]_{+}=\delta(\mathbf{p}-\mathbf{q}),
    \label{A-ex2-CAR}
\end{equation}
with all other anticommutators vanishing.

Post-selection takes place in region $\mathcal{S}$ after the field has been
turned off. Assume the postselection detector is coupled to the particle density
$\hat{O}_{s}(x) = \hat{\rho}_{\mathrm{pa}}(x)\equiv
\hat{\Phi}_{\mathrm{pa}}^{\dagger}(x)\hat{\Phi}_{\mathrm{pa}}(x)$
with
\begin{equation}
\hat{\Phi}_{\mathrm{pa}}(t,\mathbf{x})
=\int \mathrm{d} \mathbf{p}\,
\hat{b}_{\mathbf{p}}(t)u_{\mathbf{p}}(\mathbf{x})\,.
\label{A-ex2-fiplus}
\end{equation}
The postselected state is
$\ket{\chi} = p_{s}^{-1/2}\hat{K}_{\mathcal{S}}\ket{\psi}$
(see App.~\ref{app-sec:kraus}).
With $\hat{K}_{\mathcal{S}}$ given by Eq.~\eqref{kraus} of the main text
and the initial state $\ket{\psi} =\ket{0}$ being the vacuum, we have
\begin{equation}
\hat{K}_{\mathcal{S}}\ket{\psi}=
\braket{d_s|d_0}\ket{0}
-ig{\bra{d_s}\hat{P}_{s}\ket{d_0}}
\int \mathrm{d}^{1+1} x\,\Gamma_{\mathcal{S}}(x)
\hat{\rho}_{\mathrm{pa}}(x)\ket{0}\,.
\label{A-ex2-khi}%
\end{equation}
The term $\hat{\rho}_{\mathrm{pa}}(x)\ket{0}$ is computed by using
Eqs.~\eqref{A-ex2-fiplus} and \eqref{A-ex2-bogo} yielding
\begin{equation}
\hat{\rho}_{\mathrm{pa}}(x)\ket{0}
=\rho_{\mathrm{pa}}(x)\ket{0}
+\int \mathrm{d}\mathbf{p}\,\mathrm{d}\mathbf{p}^{\prime}\,
\left(  \tilde{u}_{\mathbf{p}}^{(+)}(x)\right)^{\dagger}
\tilde{w}_{\mathbf{p}^{\prime}}^{(+)}(x)\,
\hat{b}_{\mathbf{p}}^{\dagger}\hat{d}_{\mathbf{p}^{\prime}}^{\dagger}\ket{0}\,,
\label{A-ex2-rho}%
\end{equation}
where we have defined
\begin{subequations}
	\label{A-ex2-wtilde}
	\begin{align}
		\tilde{u}_{\mathbf{p}}^{(+)}(t,\mathbf{x}) &= \int\mathrm{d}\mathbf{q}\,u_{\mathbf{q}}(\mathbf{x})U_{u_{\mathbf{q}}u_{\mathbf{p}}}(t,t_{0})\,, \\
		\tilde{w}_{\mathbf{p}}^{(+)}(t,\mathbf{x}) &= \int \mathrm{d}\mathbf{q}\,w_{\mathbf{q}}(\mathbf{x})U_{u_{\mathbf{q}}w_{\mathbf{p}}}(t,t_{0})\,.
	\end{align}
\end{subequations}
These are the positive-energy components of the propagated
positive- and negative-energy Dirac basis states, respectively,
$\tilde{u}_{\mathbf{p}}^{(+)}(t,\mathbf{x})=
\bra{\mathbf{x}} \hat{U}(t,t_{0})\ket{u_{\mathbf{p}}}$ and
$\tilde{w}_{\mathbf{p}}^{(+)}(t,\mathbf{x})=
\bra{\mathbf{x}} \hat{U}(t,t_{0})\ket{w_{\mathbf{p}}}$.

The denominator of the effective weak value given by Eq.~\eqref{wv1} of the
main text is
$\braket{\chi | \psi}
=p_{s}^{-1/2}\bra{\psi} \hat{K}_{\mathcal{S}}^{\dagger} \ket{\psi}$
which is readily computed here as the vacuum expectation of the Kraus
operator,
\begin{equation}
\bra{0} \hat{K}_{\mathcal{S}}^{\dagger}\ket{0}
=\braket{d_s|d_0}^{\ast}
+ig{\bra{d_s}\hat{P}_{s}\ket{d_0}^{\ast}}
\int \mathrm{d}^{1+1} x\,\Gamma_{\mathcal{S}}(x)\rho_{\mathrm{pa}}(x)\,,
\label{A-ex2-den}%
\end{equation}
using the adjoint of Eq.~\eqref{A-ex2-khi}.
The numerator contains the term
${\bra{\chi} \hat{O}_{j}(x)\ket{\psi}}
=p_{s}^{-1/2} {\bra{\psi}\hat{K}_{\mathcal{S}}^{\dagger}\hat{O}_{j}(x)\ket{\psi}}$.
Using the adjoint of Eqs.~\eqref{A-ex2-khi} and
\eqref{A-ex2-rho}, we have
\begin{equation}
\bra{0} \hat{K}_{\mathcal{S}}^{\dagger}\hat{O}_{j}(x)\ket{0}
=\braket{d_{s} | d_{0} }^{\ast}
{\bra{0}\hat{O}_{j}(x)\ket{0}}
+ig {\bra{d_{s}} \hat{P}_{s} \ket{d_{0}}}^{\ast}
\int \mathrm{d}^{1+1} x^{\prime} \, \Gamma_{\mathcal{S}}(x^{\prime})
{\bra{0} \hat{\rho}_{\mathrm{pa}}(x^{\prime})\hat{O}_{j}(x)\ket{0}} \,,
\label{A-ex2-num1}
\end{equation}
with
\begin{equation}
\bra{0} \hat{\rho}_{\mathrm{pa}}(x^{\prime})\hat{O}_{j}(x)\ket{0}
=\rho_{\mathrm{pa}}(x^{\prime})
{\bra{0}\hat{O}_{j}(x)\ket{0}}
+\int \mathrm{d}\mathbf{p}\,\mathrm{d}\mathbf{p}^{\prime}\,
{\bra{0} \hat{d}_{\mathbf{p}^{\prime}} \hat{b}_{\mathbf{p}}\hat{O}_{j}(x)\ket{0}}
\left[  \left(  \tilde{u}_{\mathbf{p}}^{(+)}(x^{\prime})\right)^{\dagger}
\tilde{w}_{\mathbf{p}^{\prime}}^{(+)}(x^{\prime})\right]^{\ast}\,.
\label{A-ex2-num2}
\end{equation}
Plugging-in the first term of Eq.~\eqref{A-ex2-num2} into Eq.~\eqref{A-ex2-num1}
and dividing by the denominator \eqref{A-ex2-den} leaves us only with the
single term ${\bra{0} \hat{O}_{j}(x)\ket{0}}$.
The second term leads to
\begin{equation}
\frac{ \displaystyle ig{\bra{d_{s}} \hat{P}_{s}\ket{d_{0}}}^{\ast}
\int \mathrm{d}^{1+1} x^{\prime}\,\Gamma_{\mathcal{S}}(x^{\prime})
\int \mathrm{d}\mathbf{p}\,\mathrm{d}\mathbf{p}^{\prime}\,
{\bra{0} \hat{d}_{\mathbf{p}^{\prime}}\hat{b}_{\mathbf{p}}\hat{O}_{j}(x)\ket{0}}
\left[\left(  \tilde{u}_{\mathbf{p}}^{(+)}(x^{\prime})\right)^{\dagger}
\tilde{w}_{\mathbf{p}^{\prime}}^{(+)}(x^{\prime})\right]^{\ast}}
{\displaystyle \braket{d_s|d_0}^{\ast}
+ig{\bra{d_s}\hat{P}_{s}\ket{d_{0}}}^{\ast}
\int \mathrm{d}^{1+1} x^{\prime}\, \Gamma_{\mathcal{S}}(x^{\prime})
\rho_{\mathrm{pa}}(x^{\prime})}\,,
\end{equation}
which when smeared over the weak coupling region $\mathcal{W}_j$ with $\Lambda_{\mathcal{W}_j}$ gives the $C_{j,\mathcal{S}}$ term
of Eq.~\eqref{rho-weak} of the main text.

Note that for the specific choice
$\hat{O}_{j}(x)=\hat{\rho}_{\mathrm{pa}}(x)$
the matrix element entering the non-factorizable contribution can be evaluated explicitly.
Using Eqs.~\eqref{A-ex2-fiplus} and \eqref{A-ex2-bogo}, one finds
\begin{equation}
\bra{0} \hat{d}_{\mathbf{p}^{\prime}}\hat{b}_{\mathbf{p}}
\hat{\Phi}_{\mathrm{pa}}^{\dagger}(x)\hat{\Phi}_{\mathrm{pa}}(x)\ket{0}
=\left(\tilde{u}_{\mathbf{p}}^{(+)}(x)\right)^\dagger
\tilde{w}_{\mathbf{p}^{\prime}}^{(+)}(x)\,.
\end{equation}
Hence, for particle-density monitoring,
$C_{j,\mathcal{S}}$ is expressed in in Eq.~\eqref{rho-weak_Tkf} directly in terms of
the positive-energy components evolved from initial positive- and negative-energy modes.

\section{Sensing entanglement}

We derive in this Appendix the pointer shifts for the entangled and mixed
states
\begin{equation}
\rho_{\mathrm{ent}}={\ket{\psi_{\mathrm{ent}}}\bra{\psi_{\mathrm{ent}}}}\,,
\quad
\ket{\psi_{\mathrm{ent}}}
=c_{1} \ket{\varphi_{L}\eta_{R}}
+c_{2} \ket{\eta_{L}\varphi_{R}}
\label{enti-st}%
\end{equation}
and
\begin{equation}
\rho_{\mathrm{mix}}=
|c_{1}|^{2} {\ket{\varphi_{L}\eta_{R}} \bra{\varphi_{L}\eta_{R}}}
+|c_{2}|^{2} {\ket{\eta_{L}\varphi_{R}} \bra{\eta_{L}\varphi_{R}}}
\label{mixed-st}%
\end{equation}
considered in the main text.

\subsection{Kraus operator}
\label{app-sec:ex3-kraus}

Following the scheme represented in Fig.~\ref{fig:example3} we have two detectors coupling to
the field in the regions $\mathcal{S}_{L}$ and $\mathcal{S}_{R}$ defined on $\Sigma_{S}$. Given
that the pointer variable of each detector as well as the number density in
each region commute, the unitary operator $\hat{U}_{\mathcal{S}}$ describing the coupling
between the field and the detectors in $\mathcal{S}_{L}$ and $\mathcal{S}_{R}$ commute, and the
full Kraus operator factorizes:%
\begin{equation}
\hat{K}_{\mathcal{S}}=\left\langle d_{s}^{L}\right\vert \left\langle d_{s}^{R}\right\vert
\hat{U}_{\mathcal{S}}\left\vert d_{0}^{L}\right\rangle \left\vert d_{0}^{R}\right\rangle
=\left\langle d_{s}^{L}\right\vert \hat{U}_{\mathcal{S}_{L}}\left\vert d_{0}^{L}\right\rangle
\left\langle d_{s}^{R}\right\vert \hat{U}_{\mathcal{S}_{R}}\left\vert d_{0}^{R}\right\rangle
=\hat{K}_{\mathcal{S}_{L}}\hat{K}_{\mathcal{S}_{R}}\,.
\label{kfull}%
\end{equation}

Although the postselection coupling is not weak, we follow common usage and
develop each Kraus operator to first order. In the notation of
Eq.~\eqref{kraus} of the main text, we have for $\hat{K}_{\mathcal{S}}^{\dagger}$%
\begin{equation}
\hat{K}_{\mathcal{S}}^{\dagger}=\alpha_{L}^{\ast}\alpha_{R}^{\ast}I+\beta_{L}^{\ast}\alpha_{R}^{\ast}%
\hat{\rho}_{\mathcal{S}_{L}}+\alpha_{L}^{\ast}\beta_{R}^{\ast}\hat{\rho}_{\mathcal{S}_{R}}+\beta_{L}^{\ast}\beta_{R}^{\ast}%
\hat{\rho}_{\mathcal{S}_{L}}\hat{\rho}_{\mathcal{S}_{R}}\,
\label{kl-kr}%
\end{equation}
where $\hat{\rho}_{\mathcal{S}_{l}}$ is short for the smeared number density operator
$\hat{\rho}_{\mathcal{S}_{l}} \equiv
=\int\mathrm{d}^{n+1}x\,%
\Gamma_{\mathcal{S}_{l}}(x)\hat{\rho}(x)$
with $l=L,R$. We assume, as in the other examples, a
short interaction duration, or employing averaged quantities so as to use an
\textquotedblleft instantaneous\textquotedblright\ interaction
so that each interaction is effectively supported on an equal-time slice $t_{s}$.

\subsection{Effective weak value expression for a density operator}
\label{app-sec:ex3-WV}

Let the field be preselected in a state described by the general density operator $\rho$
and the full initial state be
\begin{equation}
\rho_{0}=\rho\otimes {\ket{\zeta_{0}}\bra{\zeta_{0}}}\otimes {\ket{d_{0}} \bra{d_{0}}}\,,
\label{rhototo-1}%
\end{equation}
where $\ket{\zeta_{0}}$ and $\ket{d_{0}}$ are the initial pure states of the weak pointer and postselection detector, respectively.
The weak couplings are
represented by the unitaries $\hat{U}_{j}$ as expressed in Eq.~\eqref{eq:unitary-evolution}, with $j=L,R$.
The detector coupling involves coupling a similar unitary $\hat{U}_{\mathcal{S}}$ and then projecting to the
postselected detector state $\ket{d_{s}}$. After this projection a weak pointer observable
$\hat{R}$ is readout and has the expectation value
\begin{equation}
\langle \hat{R}\rangle=\frac{\operatorname{Tr}\left(
\rho_{0}\hat{U}_{L}^{\dagger}\hat{U}_{R}^{\dagger}\hat{U}_{S}^{\dagger
}\hat{R}\left\vert d_{s}\right\rangle \left\langle d_{s}\right\vert \hat{U}_{S}%
\hat{U}_{R}\hat{U}_{L}\right)  }{\operatorname{Tr}\left(
\rho_{0}\hat{U}_{L}^{\dagger}\hat{U}_{R}^{\dagger}\hat{U}_{S}^{\dagger}\left\vert
d_{s}\right\rangle \left\langle d_{s}\right\vert \hat{U}_{S}\hat{U}_{R}
\hat{U}_{L}\right)}\,,
\label{Mtoto-1}%
\end{equation}
where, for simplicity, the pointer observable $\hat{R}$ is taken to be uniform in the region of interest:
$\hat{R}(x)=\hat{R}$ as well as $\hat{P}(x)=\hat{P}$.
With $\hat{K}_{\mathcal{S}}= {\bra{d_{s}} \hat{U}_{\mathcal{S}} \ket{d_{0}}}$ as defined in App.~\ref{app-sec:kraus}
and expanding $\hat{U}_{j}$ to first order,
the numerator of Eq.~\eqref{Mtoto-1} becomes
\begin{equation}
\operatorname{Tr} \left(  \rho {\ket{\zeta_{0}} \bra{\zeta_{0}}}
{\hat{K}_{\mathcal{S}}^{\dagger}\hat{K}_{\mathcal{S}}^{}} \hat{R}\right)
+i\lambda\sum_{j=L,R}\int \mathrm{d}^{n+1}x\,\Lambda_{\mathcal{W}_j}(x)
\operatorname{Tr}\left(  \rho_{0}\left[  \hat{O}(x)\hat{P},\,
{\hat{K}_{\mathcal{S}}^{\dagger}\hat{K}_{\mathcal{S}}^{}}\hat{R}\right]_{-}  \right)\,,
\end{equation}
and the same expression with $\hat{R}$ replaced by $I$ for the denominator.
In order to simplify further Eq.~\eqref{Mtoto-1}, pointer variables and the pointer state are assumed to obey
$\left[\hat{R},\hat{P}\right]_{-}=i$ and $\operatorname{Cov}\left(\hat{P},\hat{R}\right)=0$.
This leads to (see~\cite{jozsaComplexWeakValues2007} for a similar derivation in the case of a pure state)
\begin{equation}
\braket{\hat{R}}= \braket{\hat{R}}_{0}
+\lambda\sum_{j=L,R}\operatorname{Re}\left(O_{j}^{w}\right)\,,
\label{mean-pointer-mixed}
\end{equation}
where $\braket{\hat{R}}_{0}$ is the expectation value of the pointer observable in the state $\ket{\zeta_0}$, and
the effective weak value $O_{j}^{w}$ due to the weak coupling in region $j=L,R$ is now identified as
\begin{equation}
	O_{j}^{w}=\int \mathrm{d}^{n+1}x\,\Lambda_{\mathcal{W}_j}(x)
	\frac{\operatorname{Tr}_{\Phi}\left(  \rho {\hat{K}_{\mathcal{S}}^{\dagger}\hat{K}_{\mathcal{S}}^{}} \hat{O}_{j}(x)\right)}%
	{\operatorname{Tr}_{\Phi}(\rho {\hat{K}_{\mathcal{S}}^{\dagger}\hat{K}_{\mathcal{S}}^{}})}\,,
\label{eff-dm}
\end{equation}
which corresponds to Eq.~\eqref{eq:density-matrix-wv} of the main text, generalizing Eq.~\eqref{wv1} for density matrices.
Note that $\hat{E}_{\mathcal{S}} = {\hat{K}_{\mathcal{S}}^{\dagger}\hat{K}_{\mathcal{S}}^{}}$ is the positive operator associated with the postselection and
the corresponding postselection probability is $\operatorname{Tr}_{\Phi}(\rho \hat{E}_{\mathcal{S}})$.

\subsection{Determination of the pointer shift}
\label{app-sec:ex3-determination-pointer-shift}

The effective weak value (Eq.~\eqref{wv1} of the main text and Eq.~\eqref{eff-dm})
is built with two types of amplitudes, $\operatorname{Tr}%
_{_{\Phi}}\left(  \rho\hat{K}_{\mathcal{S}}^{\dagger}\hat{K}_{\mathcal{S}}^{}\hat{O}_{j}(x)\right)  $ and
$\operatorname{Tr}_{_{\Phi}}(\rho\hat{K}_{\mathcal{S}}^{\dagger}\hat{K}_{\mathcal{S}}^{})$. For the entangled state
for instance, the former expression gives
\begin{equation}
\begin{split}
\operatorname{Tr}_{_{\Phi}}\left(  \rho\hat{K}_{\mathcal{S}}^{\dagger}\hat{K}_{\mathcal{S}}^{}\hat{O}_{j}(x)\right)
&  =|c_{1}|^{2}\bra{\varphi_L\eta_R}\hat{K}_{\mathcal{S}}^{\dagger}\hat{K}_{\mathcal{S}}^{}\hat{O}_{j}%
(x)\ket{\varphi_L\eta_R}
+|c_{2}|^{2}\bra{\eta_L\varphi_R}\hat{K}_{\mathcal{S}}^{\dagger}\hat{K}_{\mathcal{S}}^{}\hat{O}%
_{j}(x)\ket{\eta_L\varphi_R}\\
&\quad+c_{1}^{\ast}c_{2}\bra{\varphi_L\eta_R}\hat{K}_{\mathcal{S}}^{\dagger}\hat{K}_{\mathcal{S}}^{}\hat{O}%
_{j}(x)\ket{\eta_L\varphi_R}
+c_{2}^{\ast}c_{1}\bra{\eta_L\varphi_R}\hat{K}_{\mathcal{S}}^{\dagger}\hat{K}_{\mathcal{S}}^{}\hat{O}%
_{j}(x)\ket{\varphi_L\eta_R}\,,
\end{split}
\label{numerator-entangled}%
\end{equation}
while the latter yields
\begin{equation}
\begin{split}
\operatorname{Tr}_{_{\Phi}}(\rho\hat{K}_{\mathcal{S}}^{\dagger}\hat{K}_{\mathcal{S}}^{})
&=|c_{1}|^{2}
\bra{\varphi_L\eta_R}\hat{K}_{\mathcal{S}}^{\dagger}\hat{K}_{\mathcal{S}}^{}\ket{\varphi_L\eta_R}
+|c_{2}|^{2}
\bra{\eta_L\varphi_R}\hat{K}_{\mathcal{S}}^{\dagger}\hat{K}_{\mathcal{S}}^{}\ket{\eta_L\varphi_R}\\
&\quad+c_{1}^{\ast}c_{2}
\bra{\varphi_L\eta_R}\hat{K}_{\mathcal{S}}^{\dagger}\hat{K}_{\mathcal{S}}^{}\ket{\eta_L\varphi_R}
+c_{2}^{\ast}c_{1}\bra{\eta_L\varphi_R}\hat{K}_{\mathcal{S}}^{\dagger}\hat{K}_{\mathcal{S}}^{}\ket{\varphi_L\eta_R}\,.
\end{split}
\end{equation}
For the classical mixture $\rho_{\mathrm{mix}}$ we have the same expressions
except the nondiagonal terms vanish. This leads to different pointer shifts in
the entangled and mixture cases, provided the nondiagonal elements are non-zero.

To see these nondiagonal elements are generically nonzero, compute for
instance%
\begin{equation}
\begin{split}
\bra{\varphi_L\eta_R}\hat{K}_{\mathcal{S}}^{\dagger}\hat{K}_{\mathcal{S}}^{}\hat{O}_{j}(x)\ket{\eta_L\varphi_R}
&=\bra{\varphi_L\eta_R}\Big[|\alpha_{L}|^{2}I+(\alpha_{L}^{\ast}\beta
_{L}+\beta_{L}^{\ast}\alpha_{L})\hat{\rho}_{\mathcal{S}_{L}}+|\beta_{L}|^{2}\hat{\rho}_{\mathcal{S}_{L}}^{2}\Big]\\
&  \quad\times\Big[|\alpha_{R}|^{2}I+(\alpha_{R}^{\ast}\beta_{R}+\beta
_{R}^{\ast}\alpha_{R})\hat{\rho}_{\mathcal{S}_{R}}+|\beta_{R}|^{2}\hat{\rho}_{\mathcal{S}_{R}%
}^{2}\Big]\hat{O}_{j}(x)\ket{\eta_L\varphi_R}%
\end{split}
\label{offdiag-num}
\end{equation}
appearing in $\operatorname{Tr}_{_{\Phi}}\left(  \rho\hat{K}_{\mathcal{S}}^{\dagger}\hat{K}_{\mathcal{S}}^{}\hat
{O}_{j}(x)\right)  $ or the similar term
$\bra{\varphi_L\eta_R}\hat{K}_{\mathcal{S}}^{\dagger}\hat{K}_{\mathcal{S}}^{}\ket{\eta_L\varphi_R}$ appearing in
$\operatorname{Tr}_{_{\Phi}}(\rho\hat{K}_{\mathcal{S}}^{\dagger}\hat{K}_{\mathcal{S}}^{})$. None of these quantities
is generically vanishing (an explicit calculation can be done by choosing the
wavepacket profiles, e.g.\ Gaussians, and the smearing functions). The conclusion
is that the presence of these non-diagonal terms lead to pointer statistics
that differ for the states given by Eq.~\eqref{enti-st} and \eqref{mixed-st}. \fi

\end{document}